\documentclass[a4paper]{jpconf}
\usepackage{amssymb}
\usepackage{amsthm}
\usepackage{latexsym}
\usepackage{graphicx,epsfig}
\def\be {\begin{equation}}
\def\ee {\end{equation}}
\def\ba {\begin{eqnarray}}
\def\ea {\end{eqnarray}}
\def\nn {\nonumber}
\def\bc {\begin{center}}
\def\ec {\begin{center}}

\def\a  {\alpha}

\def\m  {\mu}
\def\n  {\nu}

\def\O  {\Omega}
\def\p  {\pi}

\def\la {\label}
\def\le {\left}
\def\ri {\right}

\def\f {\frac}

\def\no {\noindent}
\def\bi {\begin{itemize}}
\def\ei {\end{itemize}}

\def\bc {\begin{center}}
\def\ec {\end{center}}

\newcommand{\bdm}{\begin{displaymath}}
\newcommand{\edm}{\end{displaymath}}

\begin{document}
\title{\textbf{
Randall-Sundrum with Kalb-Ramond field:\\
return of the hierarchy problem?
}
}
\author{\textbf{Saurya Das}}
\date{}
\address{Department of Physics, \\
University of Lethbridge \\
4401 University Drive, \\
Lethbridge, Alberta, CANADA T1K 3M4}
\ead{\textbf{saurya.das@uleth.ca }}
\author{\textbf{Anindya Dey}}
\address{Department of Physics,\\
Harish-Chandra Research Institute, \\
Chhatnag Road, Jhusi, Allahabad - 211019, India} 
\ead{\textbf{anindya@mri.ernet.in}}
\author{\textbf{Soumitra SenGupta}}
\address{Department of Theoretical Physics, \\
Indian Association for the Cultivation of Science, \\
Kolkata - 700032, India }
\ead{\textbf{tpssg@iacs.res.in}}

\begin{abstract}
We show that when the antisymmetric Kalb-Ramond field is 
included in the Randall-Sundrum scenario, 
although the hierarchy problem can be solved, it requires an
extreme fine tuning of the Kalb-Ramond field
(about $1$ part in $10^{62}$). We interpret this as the 
return of the problem in disguise.
Further, we show that the Kalb-Ramond field induces 
a small negative cosmological constant on the 
visible brane.
\end{abstract}

\section{Introduction}
The difference of about sixteen orders of magnitude 
between the electroweak scale ($\approx 1~TeV$) and the 
Planck scale ($\approx 10^{16}~TeV$), is known as the hierarchy problem. 
While theoretically there seems to be nothing which can rule
out such a difference, it certainly seems a strange thing to be.
Figure 1 demonstrates this fact.  

\begin{figure}[h]
\begin{center}
\epsfxsize 2.20 in
\epsfbox{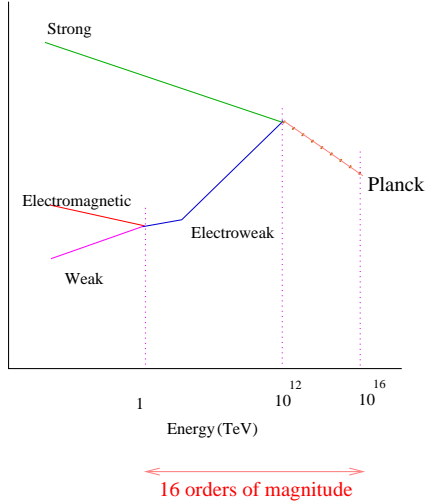}
\caption{Difference between the electroweak and Planck scale}
\end{center}
\end{figure}

Of the many attempts to explain the hierarchy problem, 
two most recent ones deserve special attention, the proposals
themselves being quite simple in themselves. 
Collectively known as the {\it Brane World Scenarios}, 
they assume the existence of one or more spatial dimensions in our
universe, in addition to the four spacetimes that we 
observe. In other words, if there are a total of $d$ spacetime
dimensions, it can be decomposed as: 
$d = {4}(Observed) + {(d-4)}(Unobserved)$, as shown diagrammatically below:
\begin{figure}[h]
\begin{center}
\epsfxsize 3.00 in
\epsfbox{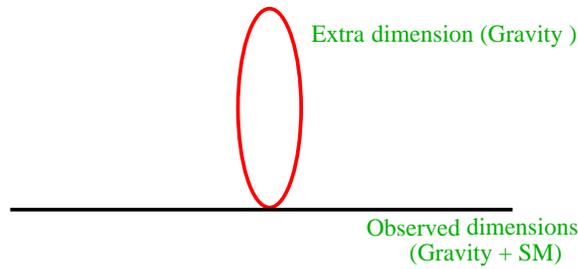}
\caption{Horizontal line = observed universe. Circle = unobserved universe.}
\end{center}
\end{figure}

\no
Here it is assumed that both Standard Model (SM) and gravity 
are present in the observable part of the universe 
(`the brane'), while gravity alone is present in the bulk. 

\subsection{ADD scanerio}

The first brane world scenario, 
known as the Arkani-Hamed-Dimopoulos-Dvali (ADD) model,
requires at least $2$ (possibly more)
extra dimensions \cite{add}. One may start with the Einstein action in 
$d$-spacetime dimensions, ${\cal R}_d$ and $G_d$ being the 
$d$-dimensional curvature scalar and gravitational constant
respectively : 
\be
S = \f{c^3}{16\p G_d} \int d^dx \sqrt{-g_d}~{\cal R}_d 
\la{ee1}
\ee
and substitute in it a $d$-dimensional metric ansatz of the
form:
\be
{ds_d}^2 = ds_4^2 - dy_I dy^I,
\la{metric1}
\ee
where the two terms represent the observed $4$-dimensional 
and the hidden $(d-4)$-dimensional parts respectively (index
$I=1,\dots,d-4$). Integrating over the unobserved dimensions,
one obtains the {\it effective} $4$-dimensional action: 
\ba
S &=& \f{c^3 V_{d-4}}{16\p G_d} \int d^4 x \sqrt{-g_4}~{\cal R}_4 \\
&\equiv& \f{c^3}{16\p G_4}  \int d^4 x \sqrt{-g_4}~{\cal R}_4
\ea
where the $4$-dimensional gravitational constant is given by:
\be
G_4 = \f{G_d}{V_{d-4}}~.
\ee
Correspondingly, the $d$-dimensional and $4$-dimensional
Planck masses are related as:
\be
M_{Pl(d)}^{d-2} = \f{\hbar^{d-3}}{c^{d-5} G_d}~~
= \f{\hbar^{d-3}}{c^{d-5} {V_{d-4}} G_4}
= \le( \f{\hbar}{c L} \ri)^{d-4} M_{Pl(4)}^2 ~,
\la{pl1}
\ee
where we have used: $V_{d-4} = L^{d-4}$. Now, the 
four dimensional (observed) Planck scale is:
\be
M_{Pl(4)} c^2 = 10^{19} GeV~=10^{16} TeV~.
\ee
Therefore, from (\ref{pl1}), the following possibilities 
(and many more) result:\\
(i) $d=6~,~L=100~\m m ~~~\Rightarrow~~M_{Pl(6)}c^2 = 1~TeV$\\
(ii) $d=10~,~L=1~Fermi ~~~\Rightarrow~~M_{Pl(10)}c^2 = 1~TeV$.
\\
In other words, the $6$ or $10$-dimensional Planck mass 
can be as low as a $TeV$. Moreover, 
these cannot be ruled, out since inverse-square law
of gravity has been tested to $0.1~mm$ so far
\footnote{for d=5, $M_{Pl(5)}c^2=1~TeV~\Rightarrow~ L>>1~mm$. Therefore
it is ruled out.} 
. 
As a result, the hierarchy problem is solved in higher
dimensions, where, more precisely, the problem ceases to exist ! 

\subsection{RS scenario}

Next, we come to the second or the Randall-Sundrum (RS) brane 
world scenario, where one again starts with the action 
(\ref{ee1}), but instead
of the metric ansatz (\ref{metric1}), one uses the following 
`warped', or non-factorisable metric \cite{rs}:
\be
ds_d^2 = e^{-A(y)} {ds_4^2} - {dy_I dy^I}~.
\la{metric2}
\ee
$\exp(-A)$ is known as the warp factor. 
Now, the effective $4$-dimensional action, the gravitational 
constant and the relation between Planck masses read as:
\ba
S &=& \f{\le[c^3 \int d^{d-4} y \sqrt{g(y)} e^{-A}\ri]}
{16\p G_d} \int d^4x \sqrt{-g_d}~{\cal R}_d \\
&\equiv & \f{c^3}{16\p G_4}  \int d^4 x \sqrt{-g_4}~{\cal R}_4 \\
G_4 &=& G_d
\le[ \int d^{d-4} y \sqrt{g(y)} e^{-A} \ri]^{-1} \\
~M_{Pl(d)}^{d-2}
&=& \le( \f{\hbar}{c} \ri)^{d-4}
\f{M_{Pl(4)}^2}{\int d^{d-4} y \sqrt{g(y)} e^{-A}} 
\approx  
 \le( \f{\hbar}{c} \ri)^{d-4}~
{k}^{d-4}~M_{Pl(4)}^2~.
\ea
In the above, a warp factor of the form $A(y)=k\sqrt{y_Iy^I}$ 
has been assumed,
which is a solution of the Einstein equations in the RS scenario,
as we shall see shortly. Now, if $d=5$, and 
our universe (`visible brane') is located at $y=y_0$, the conformal 
factor of the metric 
(\ref{metric2}) is of the form $\O^2 = e^{-A(y_0)}$.
Considering any matter action (such as that for the Higgs field) 
with a mass parameter $m_0$, and integrating 
over the extra dimension with the metric (\ref{metric2}) and the
above conformal factor results in the following {\it physical mass},
which is exponentially suppressed: 
\be
m_{H} = e^{-A(y_0)} m_{0}~.
\la{warp1}
\ee
Thus if $m_{0}^{}~c^2 = 10^{16} TeV$ and $A \approx 12$, then  
$m_{H} c^2 = 1~TeV.$ 
In other words, a small conformal factor explains hierarchy. 
The situation is depicted in the figure below, where $y\equiv r\phi$, 
$r$ being a characteristic length scale, and $y=y_0$ corresponds to
$\phi=\pi$. The warp factor on the {\it Hidden Brane} is unity.  
\begin{figure}[h]
\begin{center}
\epsfxsize 2.70 in
\epsfbox{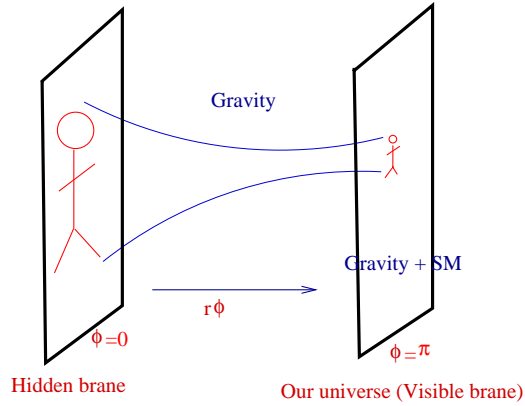}
\caption{Hidden and visible branes in the RS scenario}
\end{center}
\end{figure}

Now, we explicitly compute the warp factor in Eq.(\ref{metric2}), 
which is first written in the following form for $d=5$:
\be
ds^2=e^{-A(\phi)}~\eta_{\mu\nu}dx^{\mu}dx^{\nu} - 
r^2 d\phi^2~,
\la{metric3}
\ee
where the solution for $A$ follows from extremising the following action
($M_{Pl(5)}\equiv M$):

\begin{eqnarray}
S &=& S_{Gravity} + S_{vis} + S_{hid} \la{act1}  \\ 
{}\nn \\
\mbox{where,}~~~~S_{Gravity} &=& \int d^4x~r~d{\phi} 
\sqrt{-G}~[ 2M^3R + {\Lambda}]\\
{}\nn\\
S_{vis} &=&  \int d^4x \sqrt{-g_{vis}}~[L_{vis} - V_{vis}] \\
{}\nn\\
S_{hid} &=&  \int d^4x \sqrt{-g_{hid}}~[L_{hid} - V_{hid}]~,  
\end{eqnarray}
$\Lambda$ being the $5$-dimensional cosmological constant, and $c$ and
$G$ have been set to unity. The corresponding 
equations of motion are ($'={d}/{d\phi}$):
\begin{eqnarray}
\frac{3}{2}{A'}^2&=&-\frac{\Lambda}{4M^3}~r^2~,
\la{eom1}
\ea
which has the following solution $(k =\frac{-\Lambda}{24M^3})$ 
\ba
A&=& 2kr\phi \\
V_{hid}&=&-V_{vis}= 24 M^3 k ~.
\ea
Thus from (\ref{warp1}), we see that the following 
suppression of mass occurs:
\ba
\le( \f{m_H}{m_0}\ri)^2 = 
e^{-2A}|_{\phi=\pi} 
= e^{-2kr\p}  \approx (10^{-16})^2~, 
\la{warp2}
\ea
from which, it follows that:
\be
kr= \f{16}{\p}\ln(10)= 11.6279\dots
\ee
We will call this the `RS value' of the warp factor.

At this point, it is natural to ask as to what happens to the above value
if there are other fiels in the bulk.  
In particular, one can consider 
the massless NS-NS fields in string theory, which can
be written as
$
\a_{-1}^\mu~\tilde\a_{-1}^\nu
|0;k\rangle $
($\alpha^\mu_{-1},\tilde\alpha^\mu_{-1}$ are annihiliation/creation operators
and $|0;k\rangle$ is the string ground state),
whose symmetric, anti-symmetric and trace parts are interpreted as the 
graviton ($g_{\m\n}$), dilaton ($\phi$) and the Kalb-Ramond ($B_{\m\n})$
fields respectively. 
The last of these, which gives rise to a $3$-form field strength 
$H_{MNL} = \partial_{[M} B_{NL]}$ will be included here and its
effect on the RS scenario studied.

\section{RS scenario with Kalb-Ramond field}

We once again start with the metric ansatz (\ref{metric3}), with the
the Kalb-Ramond (KR) action added to the action (\ref{act1}). 
\begin{eqnarray}
S &=& S_{Gravity} + S_{vis} + S_{hid} + S_{KR}  \\
S_{KR} &=& \int d^4x~r~d{\phi}\sqrt{-G}~[ -2H_{MNL} H^{MNL}]~. 
\end{eqnarray}
They give rise to the following equations of motion 
(which reduce to Eq.(\ref{eom1}) when $H_{MNL}=0$): 
\begin{eqnarray}
\frac{3}{2}{A'}^2&=&-\frac{\Lambda}{4M^3}~r^2 -
\frac{3}{2M^3}
g^{\nu\beta}g^{\lambda\gamma}H_{\phi\nu\lambda}H_{\phi\beta\gamma}
\\
\frac{3}{2}({A'}^2-{A''})&=&-\frac{\Lambda}{4M^3}r^2+\frac{\exp(-2A)}{2M^3}
~\eta^{\lambda\gamma}
[- 12\eta^{00} H_{\phi 0 \lambda}H_{\phi 0 \gamma}+3\eta^{\nu\beta} 
H_{\phi\nu\lambda}H_{\phi\beta\gamma}]
\\
\frac{3}{2}({A'}^2-{A''})&=&-\frac{\Lambda}{4M^3}r^2+\frac{\exp(-2A)}{2M^3}
~\eta^{\lambda\gamma}
[-12\eta^{ii} H_{\phi i \lambda}H_{\phi i \gamma}+3\eta^{\nu\beta}
H_{\phi\nu\lambda}H_{\phi\beta\gamma}] ~.
\end{eqnarray}
Remarkably, the above set of equations has a unique solution 
of the form:
\ba
e^{-A}&=&\frac{\sqrt{b}}{2kr}\cosh{(2kr\phi+2krc)}~ \la{warp3} \\
%
%
c &=& - \frac{1}{2kr} \tanh^{-1} \left( \frac{V_{hid} }{24 M^3 k} \right)
=  -\pi + \frac{1}{2kr} \tanh^{-1} \left( \frac{V_{vis} }{24 M^3 k} \right)~,
\la{c1}\\
B^{\mu\nu} &=& k^{\mu\nu} \f{2kr}{b} \tanh\le(2kr\phi+2krc \ri)
\ea
where the parameter $b$ is related to the KR energy density, and 
$k^{\mu\nu}$ is a constant polarisation tensor. 
Requiring $A(0)=1$ on the hidden brane leads to:
\be
\frac{2kr}{\sqrt{b}}=\cosh(2krc)~.
\ee
Note that the RS limit corresponds to: $b \rightarrow 0~,~c\rightarrow-\infty$.

The counterpart of Eq.(\ref{warp2}) is now:
\ba
 \le(\f{m_H}{m_0}\ri)^2 &=& 
e^{-2A}|_{\phi=\pi} 
=
\frac{\sqrt{b}}{2kr}\cosh\left[2kr\pi + \cosh^{-1} \frac{2kr}{\sqrt{b}}\right]
\la{warp4}\\
&=&
\left[
\cosh\left(2 kr\pi\right) - \sinh\left( 2kr\pi\right)
\sqrt{1- \frac{b}{(2kr)^2}}
~\right]
\\
&\approx& (10^{-16})^2 ~,\\
{}\nn
\ea
inverting which, we get:
\ba
b = (2kr)^2
\left[1 -
\left(  
\coth(2kr \pi) - (m_H/m_0)^2 {\rm cosech}(2kr\pi)
\right)^2 
\right]~.
\nn
\ea
Note that, the RS value of $kr$ corresponds to $b=0$ (as expected).
For $kr>$ RS value, one gets $b>0$, whereas  
$kr<$ RS value corresponds to $b<0$. 
The last possibility is unphysical however, since it corresponds
to an imaginary metric and warp-factor, as can be seen from
Eqs.(\ref{warp3}) and (\ref{warp4}). 
Now let us examine the range of (positive) values of $b$, which
solve the hierarchy problem in this case. Figure 4 shows the
plot of $\log|b|$ vs $kr$. The kink corresponds to the RS value
of $kr$, for which $b=0$. The LHS of the kink corresponds to 
$b<0$ (unphysical sector), whereas the RHS corresponds to 
$b>0$ (physical sector). Note that $b$ rises to a maximum of 
$\approx 10^{-62}$ at $kr \approx 11.8$ and then falls back to zero. 
Thus, we see that although $b$ has to be non-zero, for any 
finite value of $kr$, it is extremely fine-tuned. 
It is interesting to note that such a small value of the
KR field was also predicted in a somewhat different context in
\cite{ssg1}. The hierarchy
problems appears to come back in disguise. This is our main result,
which first appeared in \cite{paper1}.

\begin{figure}[h]
\begin{center}
\epsfxsize 2.70 in
\epsfbox{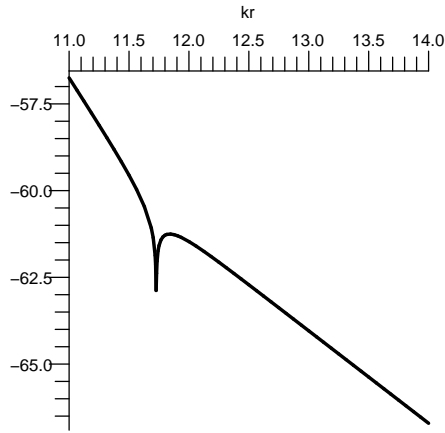}
\caption{Plot of $\log|b|$ vs $kr$}
\end{center}
\end{figure}

Next, we compute the induced $4$-dim cosmological constant on
the visible brane, which is given by \cite{maar}: 
%
%
%
%
%
%
\begin{eqnarray}
\lambda = \frac{1}{2} \left( k V_{vis} + \Lambda \right)~.
\end{eqnarray}
Using (\ref{c1}), we get:
\begin{eqnarray}
\lambda = -12 M^3 k \left[
\tanh(2krc ) + 1  \right] 
\approx 
-24 M^3 k~\frac{b}{(4kr)^2} ~ \approx -10^{-63} 
\end{eqnarray}
where in the last step, we have used the (small) value of
$b$ derived earlier. This is contrary to the currently accepted 
value of about $\lambda=+10^{-123}$ in Planck units.

\section{Summary and discussions:}

In this article, we have shown that on inclusion of the 
anti-symmetric Kalb-Ramond field, the RS brane world scenario
continues to provide a solution to the hierarchy problem, albeit with an 
extremely fine-tuned value of the KR field. In our opinion, this 
can be interpreted as the re-appearence of the problem in another
guise. Furthermore, the KR field induces a (small) negative 
cosmological constant in the visible universe, which is 
in variance with the currently accepted (small) positive 
value of the cosmological constant. 

It would be interesting to probe further phenomenological 
implications of the inclusion of the KR field, as well as those
of the dilaton field. It would also be interesting to see
whether brane worlds are stabilised against perturbations 
when these fields are present \cite{gw}. While we hope to report on these
issues elsewhere \cite{ddsg2}, here we certainly seem to be faced with the
general question: if RS brane world the answer to the hierarchy problem?

%

\ack 
SD thanks A. Dasgupta for discussions. SD and AD thank 
the Indian Association for the Cultivation of Science for hospitality, where 
part of the work was done. The work of SD
was supported by the Natural 
Sciences and Engineering Research Council of Canada, the 
University of Lethbridge and the Perimeter Institute for
Theoretical Physics.

\end{document}